\documentclass[%
 preprint,
nofootinbib,
 amsmath,amssymb,
 aps,
 prl,
 longbibliography,
 lengthcheck,%
]{revtex4-1}

\usepackage{graphicx}
\usepackage{dcolumn}
\usepackage{bm}
\usepackage{hyperref}
\usepackage{mathtools}
\usepackage[mathlines]{lineno}

\newcommand{\tr}{{\rm tr}}

\newcommand{\be}{\begin{equation}}
\newcommand{\ee}{\end{equation}}

\DeclareMathAlphabet{\mathcalligra}{T1}{calligra}{m}{n}

\begin{document}


\title{
Entangling problem Hamiltonian for adiabatic quantum computation
}

\author{Oleg Lychkovskiy$^{1,2}$}

\affiliation{$^1$ Skolkovo Institute of Science and Technology,
Skolkovo Innovation Center 3, Moscow  143026, Russia,}
\affiliation{$^2$ Department of Mathematical Methods for Quantum Technologies, Steklov Mathematical Institute of Russian Academy of Sciences,
Gubkina str. 8, Moscow 119991, Russia.}

\date{\today}


\begin{abstract}

Adiabatic quantum computation starts from embedding a computational problem into a Hamiltonian whose ground state encodes the solution to the problem. This {\it problem Hamiltonian}, $H_{\rm p}$, has been normally chosen to be diagonal in the computational basis, that is a  product basis for qubits. We point out that  $H_{\rm p}$ can be chosen to be non-diagonal in the computational basis. To be more precise, we show how to construct  $H_{\rm p}$ in such a way that all its excited states are entangled with respect to the qubit tensor product structure, while the ground state is still of the product form and encodes the solution to the problem. We discuss how such entangling problem Hamiltonians can improve the performance of the adiabatic quantum computation.

\end{abstract}

\maketitle


\noindent{\it Introduction.}  Quantum computation \cite{manin1980,benioff1980computer,benioff1982quantum,feynman1982simulating,deutsch1985quantum} promises to tackle hard computational problems inaccessible to classical computers \cite{feynman1982simulating,montanaro2016quantum}. Various models of quantum computation has been proposed to date. One such model is the adiabatic quantum computation (AQC) \cite{farhi2000quantum,farhi2001quantum}. It attracts an unceasing attention due to its elegance, implementation prospects and multiple interrelations with condensed matter physics.

Adiabatic quantum computation is based on two main ideas. The first one is that a solution of a hard computational problem can be encoded in the ground state of a quantum Hamiltonian, $H_{\rm p}$ \cite{apolloni1989quantum}, referred to as a {\it problem Hamiltonian} in what follows.
This means that there exists a mapping from the set of problem inputs (or instances) to a set of Hamiltonians, and an appropriate measurement of the ground state reveals the solution for a given input.

The second idea is that the ground state of $H_{\rm p}$ can be obtained  from a known and easily preparable ground state of another Hamiltonian, $H_0$, by slowly transforming $H_0$ to $H_{\rm p}$ in some physical device (e.g. by varying external magnetic and electric fields) \cite{farhi2000quantum,farhi2001quantum}. This physical process is described by a time-dependent Hamiltonian $H(t/T)$ with $t\in[0,T]$, which interpolates between $H_0$ and $H_{\rm p}$:
\be
H(0)=H_0, ~~~~~H(1)=H_{\rm p}.
\ee
If the run time $T$ of the computation is large enough and the ground state of $H(s)$ (where $s\equiv t/T\in[0,1]$) is nondegenerate, then, according to the adiabatic theorem \cite{born1926,born1928beweis,kato1950}, a system initiated in the ground state of $H_0$ will end up in the ground state of $H_{\rm p}$. Determining the run time  is, in general, not an easy task. For most known adiabatic algorithms the run times are not rigorously known \cite{albash2018adiabatic}.

It is well-known, however, that the run time of AQC can dramatically depend on the choices of the initial Hamiltonian $H_0$, the problem Hamiltonian $H_{\rm p}$ and the interpolating Hamiltonian $H(s)$ (here $s\equiv t/T$) \cite{albash2018adiabatic}. It is quite clear that the concept of AQC allows for a large freedom in choosing $H_0$ and $H(s)$, and wise choices are known to improve the performance of AQC \cite{albash2018adiabatic,hauke2019perspectives,albash2019role}. It is also known that for a given computational problem various different $H_{\rm p}$ can exist, and some of them are better than others \cite{choi2010adiabatic,choi2011different,choi2011differentQIC,dickson2011does,dickson2010elimination}. However, the problem Hamiltonians $H_{\rm p}$ routinely considered in the AQC studies belong to a quite narrow class of Hamiltonians. These are Hamiltonians diagonal in the computational basis, i.e. the basis constructed of product qubit states in which the final measurement is performed. It can be argued that there is a potential flaw in such choice of $H_{\rm p}$, since it forces the system to pass through  a many-body localized (MBL) or a glassy phase, leading to exponential slowdowns \cite{santoro2002theory,altshuler2010anderson,knysh2010relevance,farhi2012performance,laumann2015quantum,knysh2016zero}.  In the present paper we point out that $H_{\rm p}$ should not be necessarily diagonal in the computational basis. We show how to construct $H_{\rm p}$ with  all excited states being entangled with respect to this basis.

The reminder of the paper is organised as follows. We start from illustrating our idea with a specific example of an $NP$-complete  computational problem. Next we describe how this idea can be implemented in a general case. Last, we discuss why an entangling $H_{\rm p}$ may prove useful in evading MBL/glassy bottlenecks of the AQC.

\bigskip
\noindent{\it Monotone not-all-equal 3-satisfiability (MNAE3SAT).} This is the title of the following $NP$-complete problem. Consider a string $z=(z_1,z_2,...,z_N)$ of $N$ bits. Since the only essential property of a bit is that it is a binary variable, we are free to choose our bits to admit values $\pm 1$. An instance of a problem is a set $\cal C$ of $M$ clauses, each clause being a triple $(i,j,m)$ of pairwise nonequal integers in the interval $[1,N]$. A clause is said to be satisfied if the corresponding bits are not all equal, i.e. whenever $(z_i,z_j,z_k)\neq(1,1,1),(-1,-1,-1)$.  A solution of the problem (also called a satisfying assignment) is a bit string $z$ which satisfies all clauses from   $\cal C$. Obviously, the satisfying assumptions come in pairs related by the simultaneous flip of all bits. A discussion of this problem in the context of AQC can be found in \cite{smelyanskiy2004quantum}.

It is easy to see that MNAE3SAT is equivalent to a binary optimization problem with the cost function
\be\label{Hcl}
H^{\rm cl}_{\rm p}(z) = \sum_{(i,j,m) \in{\cal C}} C^{\rm cl}_{ijm}(z),
\ee
where
\be\label{Ccl}
C^{\rm cl}_{ijm}(z)=\left\{
\begin{array}{ll}
  1 & {\rm if}~~  z_i=z_j=z_k,\\
 0 & {\rm otherwise}.
\end{array}
\right.
\ee
If a satisfying assignment exists, then it minimizes $H^{\rm cl}_{\rm p}$, and the minimal value of $H^{\rm cl}_{\rm p}$ is zero. Vice versa, if $H^{\rm cl}_{\rm p}(z)=0$, then $z$ is a satisfying assignment.  If no satisfying assignment exist, then $H^{\rm cl}_{\rm p}(z)>0$ for any $z$. In short, to solve the problem, one has to minimize  $H^{\rm cl}_{\rm p}(z)$.

\bigskip
\noindent{\it Diagonal problem Hamiltonian.} A conventional way to map a classical binary optimization problem of the form~\eqref{Hcl} to the problem of finding the ground state of a quantum Hamiltonian is as follows \cite{albash2018adiabatic}. One considers $N$ qubits and introduces the following (formally quantum) Hamiltonian:
\be\label{H}
H_{\rm p} = \sum_{(i,j,m) \in{\cal C}} C_{ijm},
\ee
where
\be\label{C}
C_{ijm}=\frac14\left(1+\sigma_i^z\sigma_j^z+\sigma_j^z\sigma_m^z+\sigma_m^z\sigma_i^z\right),
\ee
and $\sigma_i^z$ is the third Pauli matrix for the $i$'th qubit. Note that $C_{ijm}$ is positive semi-definite.  This Hamiltonian is diagonal in the computational basis, i.e. in the common eigenbasis of all $\sigma_j^z$, $j=1,2,...,N$. Furthermore, this Hamiltonian is non-negative. Clearly, the minimization of the classical cost function \eqref{Hcl} is equivalent to finding the ground state of the Hamiltonian~\eqref{H}.
Indeed, if $z$ minimizes $H^{\rm cl}_{\rm p}$, then the product state
\be
|z\rangle\equiv |z_1,z_2,...,z_N\rangle,~~~~~~\sigma_j^z |z\rangle=z_j |z\rangle
\ee
is the ground state of $H_{\rm p}$, and vice versa.


\bigskip
\noindent{\it Entangling problem Hamiltonian.} To summarize the previous section, a ground state of the Hamiltonian \eqref{H} with a particular set of clauses $\cal C$ encodes a solution of a particular instance of the MNAE3SAT problem. This ground state is a product state $|z\rangle$, which allows one to reveal the solution $z$ by a series of $N$ single-qubit measurements. Note, however, that all other eigenstates of $H_{\rm p}$ are also product states, up to degeneracies. This latter feature is absolutely unnecessary for purposes of computation. Furthermore, it is likely to be even harmful, as discussed in what follows.

We point out that one can easily avoid this feature by introducing another problem Hamiltonian,
\be\label{entangling H}
H^{\rm ent}_{\rm p} =  \sum_{(i,j,m) \in{\cal C}} C_{ijm} A_{ijm}C_{ijm}.
\ee
Here operators $A_{ijm}$ are arbitrary self-adjoint positive-definite operators. In particular, $A_{ijm}$ can act nontrivially on qubits other than $i$'th, $j$'th and $m$'th qubits (indexes $i,j,m$ in $A_{ijm}$ indicate nothing more that $A_{ijm}$ is sandwiched between two operators  $C_{ijm}$). Further, operators $A_{ijm}$ can act on some auxiliary degrees of freedom different from $N$ qubits required for computation.  The construction \eqref{entangling H} is the main result of the present paper.

Importantly, one can choose operators $ A_{ijm}$ which are non-diagonal in the computational basis and do not commute with $ C_{ijm}$. Given such a choice, $H^{\rm ent}_{\rm p} $ is also non-diagonal. To be more precise, excited eigenstates of $H^{\rm ent}_{\rm p} $ are generically {\it entangled} with respect to the computational basis. However, it is easy to see that if a satisfying assignment for the optimization problem \eqref{Hcl},\eqref{Ccl} exists,  then  the ground states of  $H^{\rm ent}_{\rm p}$ can be chosen to be product states and coincide with the ground states of $H_{\rm p}$, and the ground state energy of $H^{\rm ent}_{\rm p} $ is zero.\footnote{In other words,  if a satisfying assumption exists, then $H^{\rm ent}_{\rm p}$ is  frustration-free. The latter property  means \cite{bravyi2009complexity} that there exists a ground state $|z\rangle$  of $H_{\rm p}$ that is also a ground state of any  $C_{ijm}$,
\be
H^{\rm ent}_{\rm p}|z\rangle=0,\qquad\forall ~(i,j,m)\in{\cal C}~~~~C_{ijm}|z\rangle=0.
\ee
Of course, the same property trivially holds for $H_{\rm p}$. However, $H^{\rm ent}_{\rm p}$ does not, in general, commute with $C_{ijm}$, in contrast to $H_{\rm p}$. } In short, $H^{\rm ent}_{\rm p} $ is a valid problem Hamiltonian for the computational problem  \eqref{Hcl},\eqref{Ccl}.

\begin{figure*}[t] 
	\centering
	\includegraphics[width=0.4\textwidth]{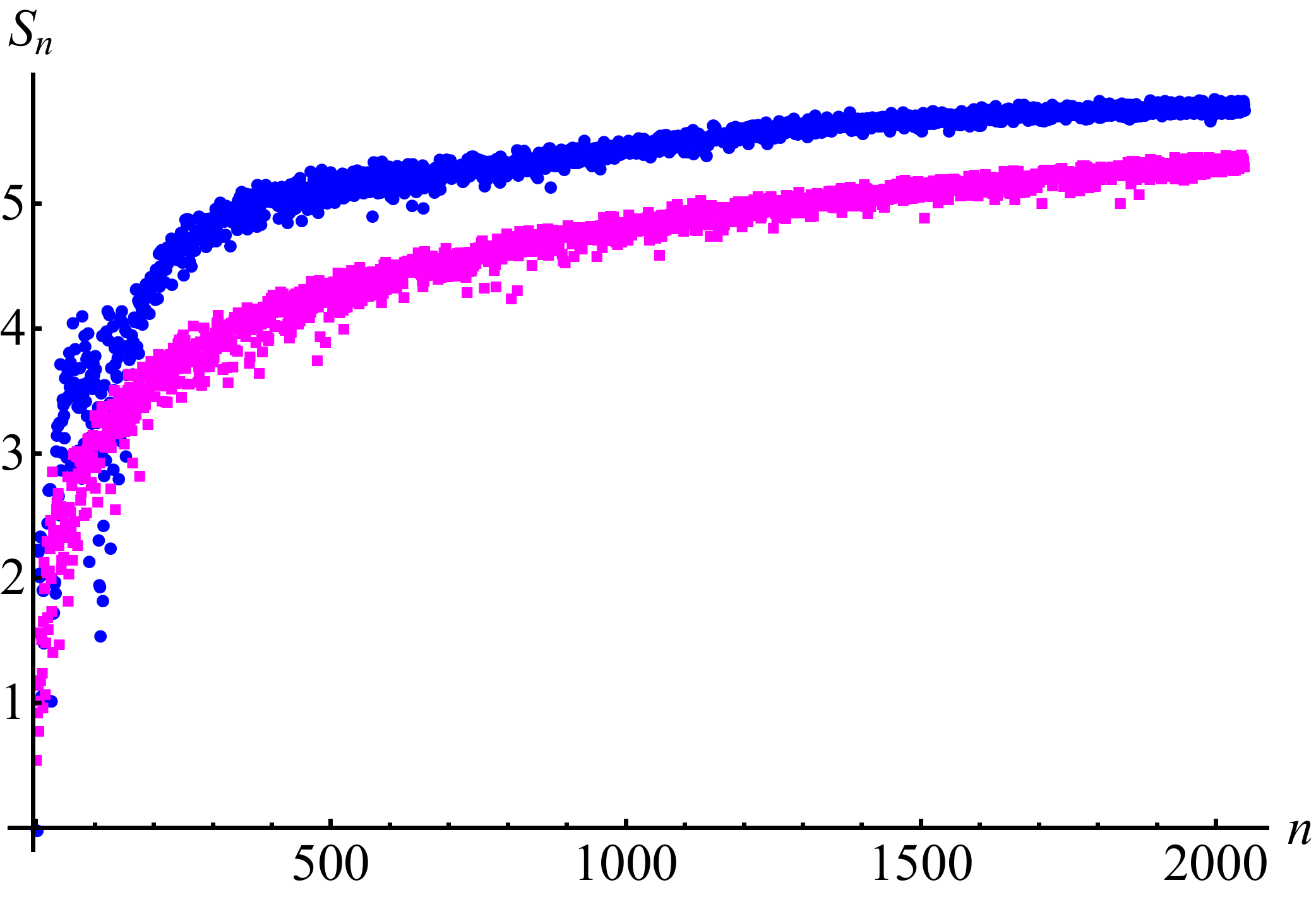}~~~~~~~~~~~~~~~~~~~~~~
	\includegraphics[width=0.4\textwidth]{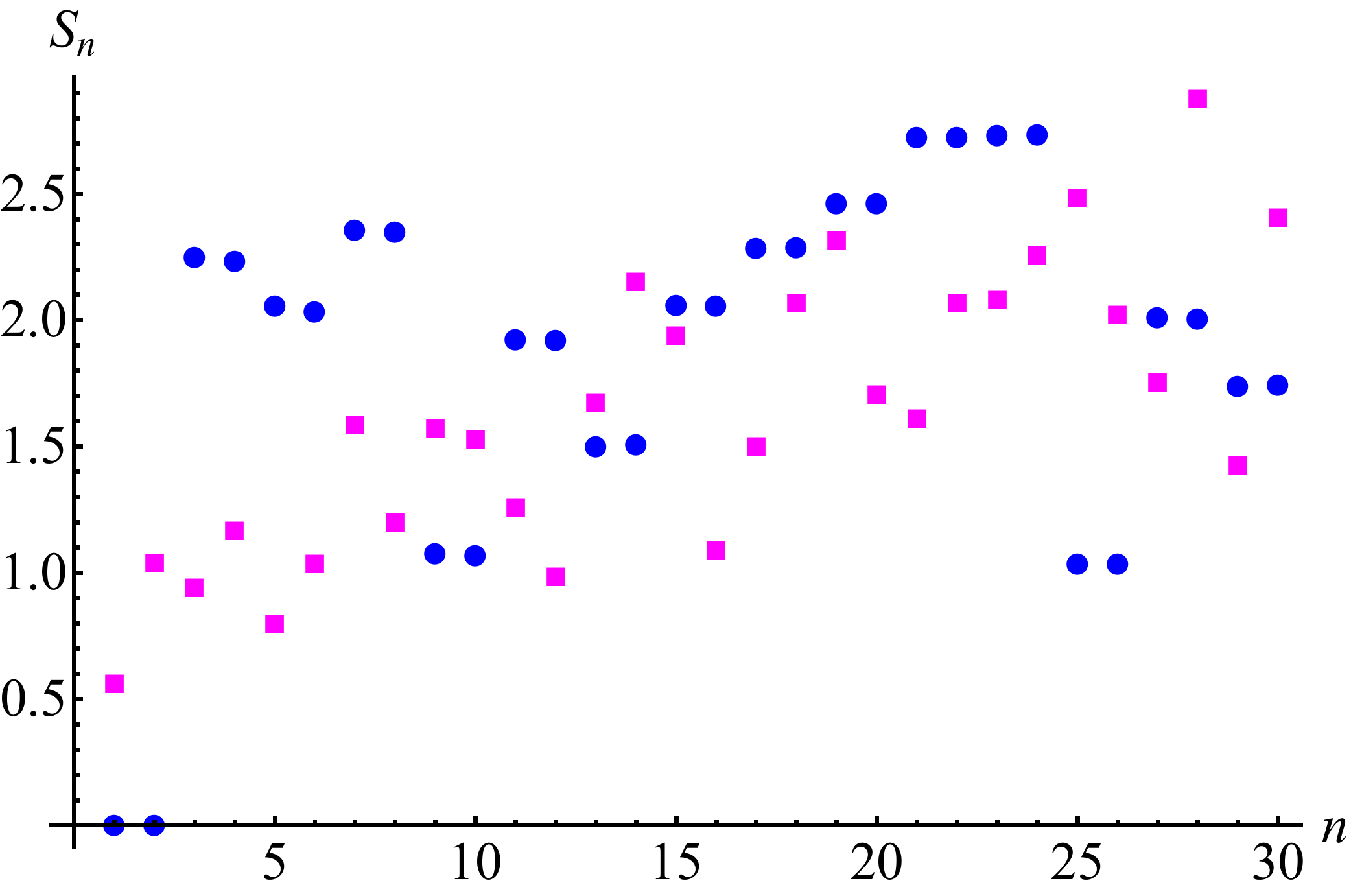}
	\caption{ Eigenstate entanglement entropy for the entangling problem Hamiltonian \eqref{entangling H} with $A_{ijm}$ given by eq. \eqref{A} (blue dots). Left plot -- the first quarter of the spectrum, right plot -- zoom to $30$ lowest eigenstates. The number of qubits is $N=14$. Eigenstates are ordered by eigenenergies, the first two being two degenerate ground states. These ground states are of product form, thus zero entanglement entropy. They  encode two satisfying assignments  of the given instance of MNAE3SAT. All excited states are entangled.  For comparison, plotted is the eigenstate entanglement entropy for a nonintegrable Ising model with $N=14$ spins $1/2$ (magenta squares).  }
	\label{fig}
\end{figure*}

Note that if $C_{ijm}$ is $k_C$-local and $A_{ijm}$ is $k_A$-local, then $H_{\rm p}$ is $k_C$-local while  $H^{\rm ent}_{\rm p}$ is, in general $(2k_C+k_A)$-local (although the locality can be tighter in certain cases). In this respect $H_{\rm p}$ has an advantage compared to $H^{\rm ent}_{\rm p}$, since tighter locality is favorable for physical implementations.

Interestingly, a construction similar to that in eq. \eqref{entangling H} was used to embed non-thermal low-entropy states into the middle of the spectrum of an otherwise chaotic many-body model and thus demonstrate the breakdown of the eigenstate thermalization hypothesis \cite{shiraishi2017systematic}.

\bigskip
\noindent{\it Specific example.} Clearly, the freedom of choice for  $ A_{ijm}$ is almost unlimited. We have studied in some detail  a Hamiltonian $H^{\rm ent}_{\rm p}$ with $A_{ijm}$ of a simple form
\be\label{A}
A_{ijm}=A=1+N^{-1}\sum_{i=1}^N \sigma_i^x.
\ee
%
%
To quantify to what extent  the excited states of $H^{\rm ent}_{\rm p}$ differ from the product states we employ the entanglement entropy $S_n=-\tr \left(\rho_n\log_2 \rho_n\right)$, where $\rho_n$ is the reduced density matrix obtained from the pure eigenstate $|n\rangle\langle n |$ of  $H^{\rm ent}_{\rm p}$ by tracing out half of the qubits (assuming their total number is even). The entanglement entropy vanishes for a product state and equals to $N/2$ for a maximally entangled state.

We  diagonalize the Hamiltonian \eqref{entangling H} with  $ A_{ijm}$ defined by eq. \eqref{A} for a small system with $N=14$ qubits and a randomly chosen set $\cal C$ of triples $(i,j,m)$.\footnote{We require that the set  $\cal C$ consists of 31 triples, and there are exactly two satisfying assignments.} In Fig. \ref{fig} we show the entanglement entropy of eigenstates of  $H^{\rm ent}_{\rm p}$. Two ground states representing the satisfying assignments of  $\cal C$ are of product form and have zero entanglement entropy. All excited states have $S_n>0$, which confirms that they are entangled. The degree of their entanglement is comparable to that for a quantum ergodic many-body system, as discussed in what follows.

\bigskip
\noindent{\it Many-body localization and $H^{\rm ent}_{\rm p}$.} Now we are in a position to discuss the relation between our construction of $H^{\rm ent}_{\rm p}$ and obstructions to AQC due to passage through the many-body localized phase \cite{santoro2002theory,altshuler2010anderson,knysh2010relevance,farhi2012performance,knysh2016zero,laumann2015quantum}. Consider a system of qubits with a Hamiltonian which is local in the computational basis. An eigenstate of a disordered Hamiltonian is said to be many-body localized if, roughly speaking, it can be expanded over a small number of states diagonal in the computational basis (see e.g. review \cite{alet2018many} and references therein). A precise definition of smallness would require a quantitative criterion in terms of scaling with the number of qubits, as well as a clear distinction between a tensor product structure and a set of bases which are of product form with respect to this tensor product structure. We do not elaborate upon such a definition here. Instead, we note that all eigenstates of the conventional problem Hamiltonian $H_{\rm p}$ given by \eqref{H} are product states and thus are, arguably, many-body localized in an ultimate manner. Thus $H_{\rm p}$ is likely to lie deep in the many-body localized phase in the parameter space of local qubit Hamiltonians~\cite{laumann2015quantum}. Therefore, the final section of the path in the parameter space corresponding to $H(s)$ inevitably lies in the MBL phase. This is believed to be accompanied by exponentially small energy gaps which lead to exponential slowdown of AQC \cite{laumann2015quantum,knysh2016zero}.

Employing $H_{\rm p}^{\rm ent}$ as the problem  Hamiltonian might mitigate this problem. Indeed, all excited states of $H_{\rm p}^{\rm ent}$ are generically entangled, and only ground states are of product form.  Our numerical experiments with small systems indicate that entanglement entropies of excited eigenstates of  $H_{\rm p}^{\rm ent}$ are quite high -- in fact, comparable to those of a bona-fide ergodic (and non-localized) quantum system. This is illustrated in  Fig. \ref{fig}, where we compare eigenstate entanglement entropies of  $H_{\rm p}^{\rm ent}$ and a paradigmatic ergodic system -- a non-integrable Ising model \cite{kim2013ballistic,kim2014testing}.\footnote{
The Ising model we choose for comparison is given by
$H_{\rm Ising}=0.9 \sum_{j=1}^{N}\sigma^x_j+0.8 \sum_{j=1}^{N}(1-0.3 j/N)\sigma^z_j+\sum_{j=1}^{N-1}\sigma^z_j\sigma^z_{j+1}$, which is a slight modification of the Hamiltonian studied in \cite{kim2013ballistic,kim2014testing}. We have verified that its spectrum indeed exhibits clear sines of ergodicity for the finite size system with $N=14$.
}  While further studies with larger system sizes are necessary to establish the exact ergodic/localization properties of $H_{\rm p}^{\rm ent}$,  it seems highly plausible that this Hamiltonian is  closer to ergodic phase than $H_{\rm p}$. Further work is required to establish what advances in the AQC performance can be gained by using $H_{\rm p}^{\rm ent}$ instead of $H_{\rm p}$. These advances should be weighted against the unfavorable locality properties of $H_{\rm p}^{\rm ent}$ as well as increased number of couplings, as compared to $H_{\rm p}$.

\bigskip
\noindent{\it Summary and concluding remarks.} To summarize, we have constructed a quantum Hamiltonian $H_{\rm p}^{\rm ent}$ whose ground state encodes a solution to a $NP$-complete problem. This Hamiltonian can be used as a problem Hamiltonian in adiabatic quantum computation.  The ground state of $H_{\rm p}^{\rm ent}$ is of the product form and coincides with the ground state of a conventional problem Hamiltonian,  $H_{\rm p}$. However, all excited states of $H_{\rm p}^{\rm ent}$ are entangled, in sharp contrast to those of   $H_{\rm p}$. We hope that this feature can prove useful for adiabatic quantum computation. We have provided some arguments related to many-body localization why this can be the case.

A few remarks are in order.
%
%
%
First,  the Hamiltonian \eqref{entangling H} is a particular case of a more general Hamiltonian with analogous properties,
\be\label{entangling H general}
\widetilde H^{\rm ent}_{\rm p} =  \sum_{\substack{(i,j,m) \in{\cal C}\\(n,l,q) \in{\cal C}}} C_{nlq} A_{ijm}^{nlq}C_{ijm},
\ee
where $A_{ijm}^{nlq}=A^{ijm}_{nlq}$ are arbitrary self-adjoint positive-definite operators. This generalization provides even more freedom for choosing the problem Hamiltonian for the adiabatic quantum computation.

Next, our construction is not limited to the particular computational problem considered. In fact, it applies to any computational problem equivalent to finding a satisfying assignment for a function $H^{\rm cl}_{\rm p}(z) = \sum_{\nu} C^{\rm cl}_\nu(z)$, where each $C^{\rm cl}_\nu(z)\geq 0$ and $C^{\rm cl}_\nu(z)=0$ for a satisfying assignment.

Finally, the impact of various choices of operators $A_{ijm}$ in eq. \eqref{entangling H} (or $A_{ijm}^{nlq}$ in eq. \eqref{entangling H general}) is yet to be explored. Of particular interest is the case of bosonic $A_{ijm}$, since coupling to the bosonic bath is known to destroy the many-body localization \cite{nandkishore2015many}, and one may hope to avoid the MBL bottleneck completely (see a related proposal \cite{pino2019mediator} to introduce bosonic couplings between spins). This is a promising direction for further research.



\begin{acknowledgments}
\bigskip
\noindent{\it Acknowledgements.} The author is grateful to B. Fine and V. Dobrovitski for useful comments. The work was supported by the Russian Science Foundation under the grant N$^{\rm o}$ 17-71-20158.
\end{acknowledgments}

\bibliography{C:/D/Work/QM/Bibs/LZ_and_adiabaticity,C:/D/Work/QM/Bibs/AQC,C:/D/Work/QM/Bibs/mbl,C:/D/Work/QM/Bibs/thermalization}

\end{document}